\documentclass[conference,a4paper]{APSIPA2025}
\usepackage{amsmath}
\usepackage{graphicx}
\usepackage{multirow}
\usepackage{threeparttable}
\usepackage{xcolor}
\usepackage{times}
\usepackage{latexsym}
\usepackage{booktabs}
\usepackage{verbatim}
\usepackage[table,xcdraw]{xcolor}
\usepackage{multirow}
\usepackage[table,xcdraw]{xcolor}
\usepackage{colortbl}  
\usepackage[table,xcdraw]{xcolor}
\usepackage{booktabs}
\usepackage[T1]{fontenc}
\usepackage[utf8]{inputenc}
\usepackage{microtype}
\usepackage{amsmath, amssymb, graphicx, xcolor, multirow, multicol, comment, subcaption, algorithm2e, url, float, etoolbox, adjustbox, pgf, soul, geometry, colortbl, booktabs}
\definecolor{wine20}{RGB}{190,80,90}   
\definecolor{wine15}{RGB}{210,120,130}  
\definecolor{wine10}{RGB}{230,160,170}  
\definecolor{wine5}{RGB}{245,210,210}   

\usepackage[backend=biber,style=ieee,]{biblatex}
\usepackage{geometry}
\usepackage{fancyhdr}

\fancypagestyle{firststyle}{
  \fancyhf{}
  \fancyhead[C]{2025 Asia Pacific Signal and Information Processing Association Annual Summit and Conference (APSIPA ASC)}
}

\begin{document}

\title{Are Multimodal Foundation Models All That Is Needed for Emofake Detection?}

\author{
\authorblockN{
Mohd Mujtaba Akhtar\authorrefmark{2}\authorrefmark{3}\authorrefmark{1},
Girish\authorrefmark{3}\authorrefmark{4}\authorrefmark{1},
Orchid Chetia Phukan\authorrefmark{3}\authorrefmark{1},
Swarup Ranjan Behera\authorrefmark{5}\\
Pailla Balakrishna Reddy\authorrefmark{6},
Ananda Chandra Nayak\authorrefmark{7},
Sanjib Kumar Nayak\authorrefmark{8},
Arun Balaji Buduru\authorrefmark{3}
}

\authorblockA{
\authorrefmark{2}V.B.S.P.U, India,
\authorrefmark{3}IIIT-Delhi, India,
\authorrefmark{4}UPES, India,
\authorrefmark{5}Independent Researcher, India\\ 
\authorrefmark{6}Reliance AI, India,
\authorrefmark{7}KAC, India,
\authorrefmark{8}VSSUT, India \\
E-mail: \textcolor{blue}{orchidp@iiitd.ac.in}
}
}

\maketitle

\begingroup
  \renewcommand{\thefootnote}{\fnsymbol{footnote}}
  \setcounter{footnote}{0}
  \footnotetext{* Contributed equally as first authors.}
\endgroup
\thispagestyle{firststyle}
\pagestyle{fancy}

\begin{abstract}
In this work, we investigate multimodal foundation models (MFMs) for EmoFake detection (EFD) and hypothesize that they will outperform audio foundation models (AFMs). MFMs due to their cross-modal pre-training, learns emotional patterns from multiple modalities, while AFMs rely only on audio. As such, MFMs can better recognize unnatural emotional shifts and inconsistencies in manipulated audio, making them more effective at distinguishing real from fake emotional expressions. To validate our hypothesis, we conduct a comprehensive comparative analysis of state-of-the-art (SOTA) MFMs (e.g. LanguageBind) alongside AFMs (e.g. WavLM). Our experiments confirm that MFMs surpass AFMs for EFD. Beyond individual foundation models (FMs) performance, we explore FMs fusion, motivated by findings in related research areas such synthetic speech detection and speech emotion recognition. To this end, we propose \texttt{\textbf{SCAR}}, a novel framework for effective fusion. \texttt{\textbf{SCAR}} introduces a nested cross-attention mechanism, where representations from FMs interact at two stages sequentially to refine information exchange. Additionally, a self-attention refinement module further enhances feature representations by reinforcing important cross-FM cues while suppressing noise. Through \texttt{\textbf{SCAR}} with synergistic fusion of MFMs, we achieve SOTA performance, surpassing both standalone FMs and conventional fusion approaches and previous works on EFD.
\end{abstract}

\section{Introduction}
In recent years, significant research has focused on advancing audio deepfake detection techniques to address various categories of deepfakes, including synthetic speech, scene-fake, and singfake. Synthetic speech deepfakes leverage advance text-to-speech (TTS) and voice conversion (VC) models to generate highly realistic voices, often indistinguishable from genuine recordings \cite{wang2020asvspoof}. 
Scene-fake manipulations involve altering background acoustic environments to misrepresent situational context \cite{yi2024scenefake}, while singfake techniques modify vocal attributes to synthesize high-fidelity singing voices \cite{zang2024singfake}. 
These deepfakes pose serious risks, such as identity fraud and misinformation through synthetic speech, manipulation of forensic evidence via scene-fake techniques, and copyright infringement or reputational harm from sing-fake synthesis. Such threats undermine trust in audio authenticity, highlighting the need for robust detection measures. As such researchers have proposed various techniques for audio deepfake detection \cite{wang2017spoofing, tran24_interspeech, guragain2024speech}. However, with the wider and easier availability of foundation models (FMs) in recent times, audio deepfake detection research has seen sufficient advancement \cite{guo2024audio, pascu24_interspeech, yang2024robust}. These FMs are pre-trained on diverse large-scale audio data and provide uplift in performance as well as prevention of training models from scratch and this trend of leveraging FMs is consistent across synthetic speech detection \cite{kawa23b_interspeech}, scenefake detection \cite{kheir2025comprehensive}, and singfake detection \cite{chen24o_interspeech}. Despite these significant advancements, an emerging vulnerability remains largely unaddressed i.e. emotionally manipulated speech deepfakes (EmoFake). 

\begin{figure}[hbt!]
    \centering
    \includegraphics[width=0.7\linewidth]{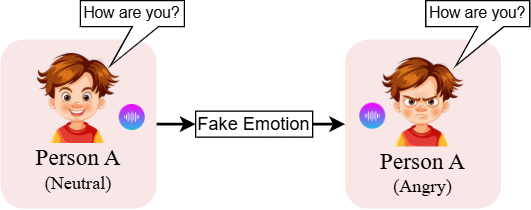}
    \caption{Demonstration of EmoFake: Speaker A's happy speech is manipulated to synthesize a sad emotional tone in the audio while maintaining the same spoken content ('I'm fine, thanks!').}
    \label{fig:emofake}
\end{figure}

EmoFake (Figure \ref{fig:emofake}) alters a speaker’s emotional attributes while preserving linguistic content and speaker identity. Unlike conventional synthetic speech, it subtly modifies pitch, tone, and intensity to create realistic emotional manipulations. This emerging threat poses serious risks in security, forensics, and misinformation, enabling malicious actors to exploit emotional deception for fraud and manipulation. As such Zhao et al. \cite{zhao2024emofake} proposed the first initial dataset for EmoFake Detection (EFD) and presented initial baselines on it. However, they haven't explored FMs for EFD which have shown its efficacy in detecting various types of audio deepfakes. In this study, we focus on EFD and explore FMs for the first time for EFD, to the best of our knowledge. \textit{We hypothesize that multimodal FMs (MFMs) such as LanguageBind (LB), ImageBind (IB) will outperform audio FMs (AFMs) (WavLM, wav2vec2, Unispeech-SAT, etc.) for EFD. Unlike AFMs, which rely solely on audio, MFMs due to their cross-modal pre-training, learns emotional patterns across multiple modalities and this strength enables MFMs to more effectively detect unnatural emotional shifts and inconsistencies in manipulated audio, making them superior at distinguishing genuine from fake emotional expressions.} To test our hypothesis, we conduct a comprehensive comparative analysis of state-of-the-art (SOTA) MFMs and AFMs. Our experiments confirm that MFMs consistently outperform AFMs in EFD. Further, inspired from research in synthetic speech detection \cite{chetia-phukan-etal-2024-heterogeneity}, speech emotion recognition \cite{wu2023investigation} where fusion of FMs have shown improved performance. We also explore this direction and we are the primary study to explore fusion of FMs for EFD. To this end, we introduce \textbf{\texttt{SCAR}} (Ne\textbf{\texttt{S}}ted \textbf{\texttt{C}}ross-\textbf{\texttt{A}}ttention Netwo\textbf{\texttt{R}}k), a novel framework for effective fusion of FMs. \textbf{\texttt{SCAR}} employs a novel nested cross-attention mechanism that enables multi-stage interaction between FMs representational space, refining information exchange. Additionally, a self-attention refinement module enhances feature representations by amplifying critical cross-FMs cues while suppressing noise. By leveraging \textbf{\texttt{SCAR}} with the fusion of MFMs, we achieve topmost performance surpassing both standalone FMs, baseline fusion techniques, and thus achieving SOTA in comparison to previous SOTA work for EFD.

\noindent \textbf{Our study offers the following key contributions}:
\begin{itemize}
\item We present the first comprehensive comparative exploration into various SOTA MFMs and AFMs to investigate the efficacy of MFMs for EFD. We show that MFMs are the best for EFD attributing to their multimodal pre-training.
\item We propose, \textbf{\texttt{SCAR}} for fusion of FMs. With \textbf{\texttt{SCAR}} through the fusion of MFMs, we report the topmost performance against individual FMs, baseline fusion techniques and setting SOTA in comparison to previous works. 
\end{itemize}

\noindent \textit{The resources i.e., models and code are accessible at \footnote{\url{https://github.com/Helix-IIIT-Delhi/SCAR-EmoFake}} for future research.}

\section{Foundation Models}
In this section, we discuss the SOTA AFMs and MFMs used in our study.

\noindent \textbf{Audio Foundation Models}: Unispeech-SAT\footnote{\url{https://huggingface.co/microsoft/unispeech-sat-base}} \cite{chen2022unispeech}, Wav2vec2\footnote{\url{https://huggingface.co/facebook/wav2vec2-base}} \cite{baevski2020wav2vec}, WavLM\footnote{\url{https://huggingface.co/microsoft/wavlm-base}} \cite{chen2022wavlm}, and HuBERT\footnote{\url{https://huggingface.co/facebook/hubert-base-ls960}} \cite{hsu2021hubert} were pre-trained in self-supervised manner. We use the base versions of Unispeech-SAT, Wav2vec2, WavLM, HuBERT with 94.68M, 95.04M, 94.70M, 94.68M parameters respectively and pre-trained on librispeech 960 hours of english data. Unispeech-SAT and WavLM reports the SOTA performance across different tasks in SUPERB. Unispeech-SAT was pre-trained in multi-task speaker-aware format and WavLM was pre-trained for solving masked speech modeling and speech denoising simultaneously. Wav2vec2 follows contrastive objective during its pre-training while HuBERT solves BERT-like mask prediction objective. We also consider Whisper\footnote{\url{https://huggingface.co/openai/whisper-base}} \cite{radford2023robust} that has shown its effectiveness for related synthetic speech deepfake detection \cite{kawa2023improved} and singfake detection \cite{10887871}. Whisper was pre-trained on 96 languages and follows vanilla transformer encoder-decoder architecture. It was pre-trained in a multi-task learning manner and we use its base version with 74M parameters. We resample the audio samples to 16KHz before feeding to the AFMs. We extract representations from the last hidden layer of the frozen AFMs by application of pooling-average. The extracted representations are of dimension: 768 for Unispeech-SAT, Wav2vec2, WavLM, HuBERT; 512 for Whisper. For Whisper, we use only its encoder for extracting representations by discarding the decoder.

\noindent \textbf{Multimodal Foundation Models}: We use IB\footnote{\url{https://github.com/facebookresearch/ImageBind/tree/main}} \cite{girdhar2023imagebind} and LB\footnote{\url{https://github.com/PKU-YuanGroup/LanguageBind}} \cite{zhu2023languagebind} for MFMs. IB is a multimodal learning framework that aligns modalities— images, text, audio, depth, thermal, and IMU—into a unified embedding space with image modality as the binding modality. It empolys a VIT enocder with contrastive learning objective. Without requiring direct supervision for all modality pairs, IB enables zero-shot recognition and retrieval for multimodal tasks. LB aligns multiple modalities such as video, audio, depth, and infrared—directly to the language space, eliminating reliance on image-based alignment. It employs a contrastive learning approach with a frozen 12-layer Transformer-based language encoder. It was pre-trained on VIDAL-10M, a dataset comprising 10 million video-text, infrared-text, depth-text, and audio-text pairs. We resample the audios to 16KHz before passing to the MFMs and extract representations from the last hidden state of the audio encoders of the MFMs by using average pooling. We extract representations of 768, 1024 size for LB and IB respectively. \newline




\begin{figure*}[hbt!]
    \centering
    \begin{subfigure}[b]{0.24\textwidth}
        \includegraphics[width=\textwidth]{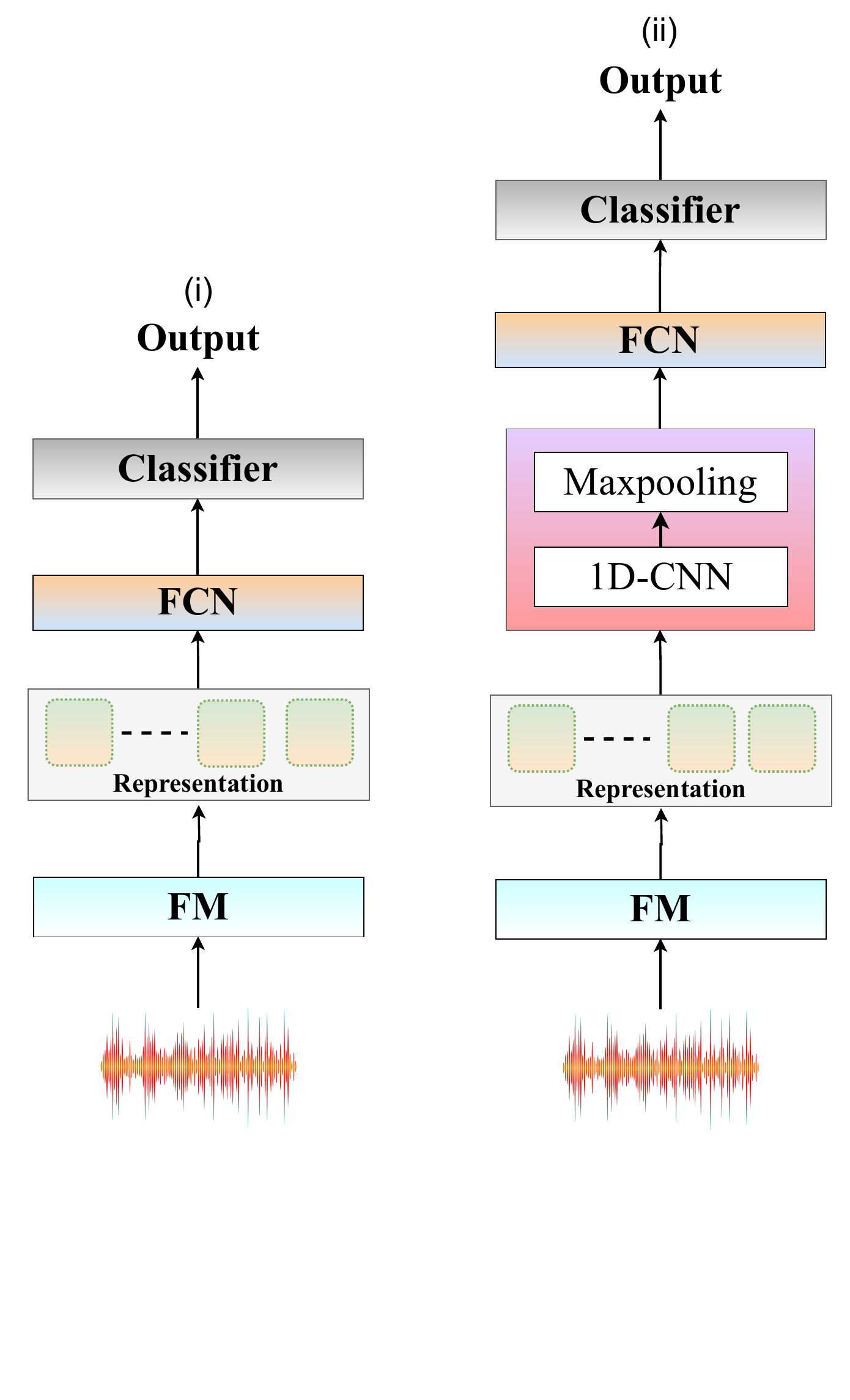}
        \caption{}
        \label{fig:image1}
    \end{subfigure}
    \hspace{0.04\textwidth} 
    \begin{subfigure}[b]{0.24\textwidth}
        \includegraphics[width=\textwidth]{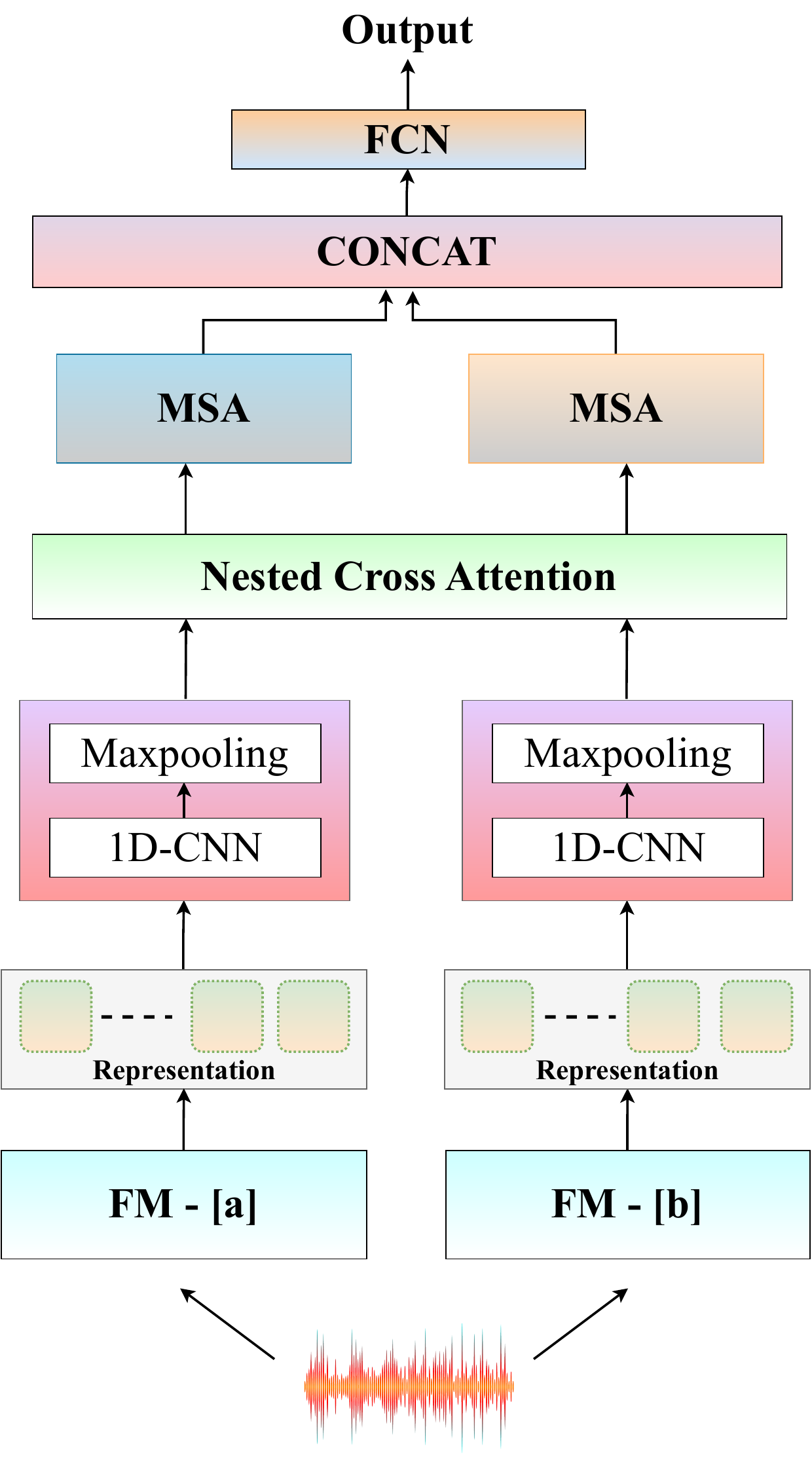}
        \caption{}
        \label{fig:image2}
    \end{subfigure}
    \hspace{0.04\textwidth} 
    \begin{subfigure}[b]{0.24\textwidth}
        \includegraphics[width=\textwidth]{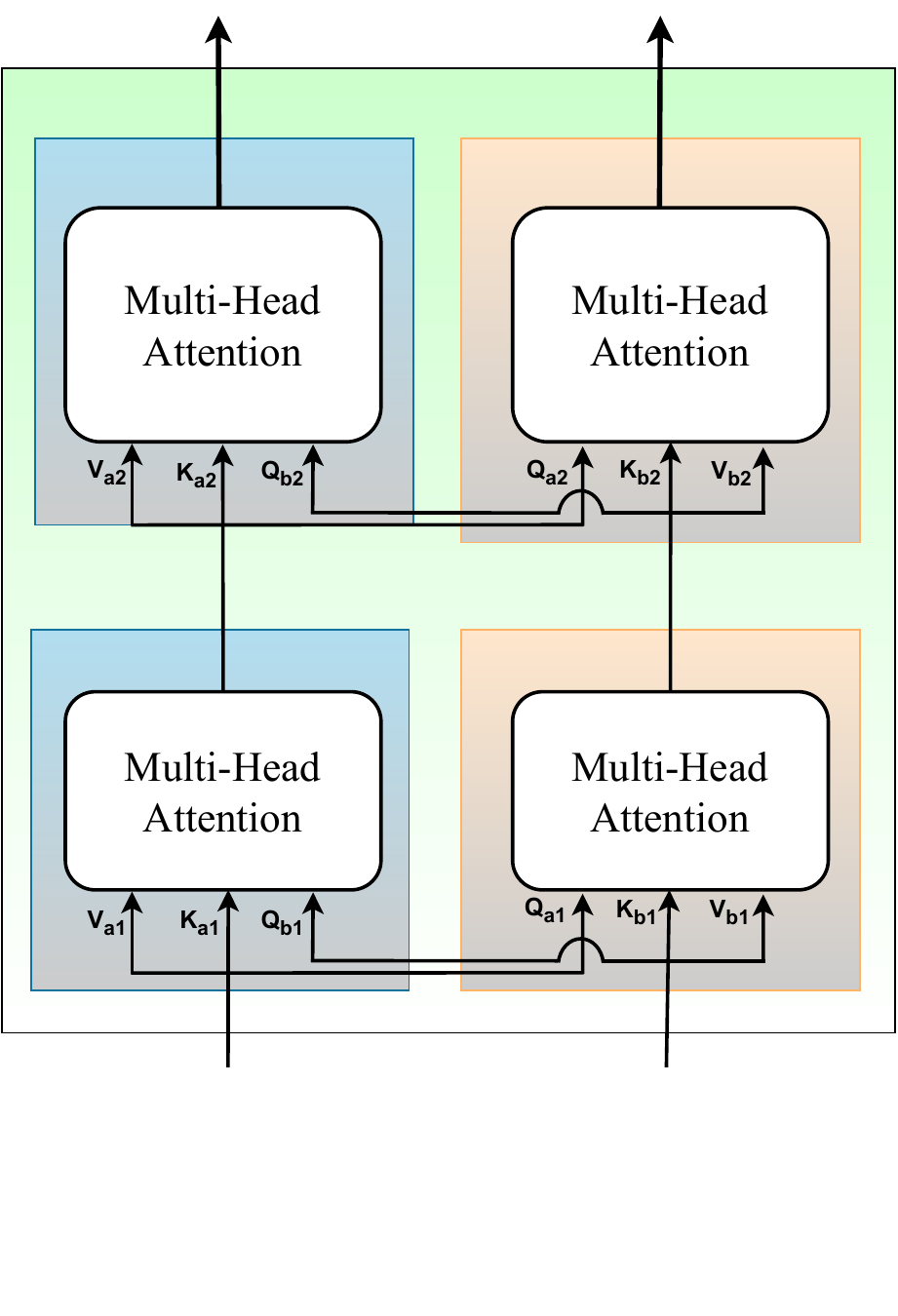}
        \caption{}
        \label{fig:image2}
    \end{subfigure}
    \caption{Modeling Architectures: Subfigure (a) FCN and CNN; Subfigure (b) shows the proposed framework, \textbf{\texttt{SCAR}}; Subfigure (c) provides the detailed illustration of the nested cross-attention mechanism}
    \label{architecture}
\end{figure*}

\section{Modeling}
In this section, we give detailed explanation of the downstream modeling involved with individual FMs followed by the proposed framework, \textbf{\texttt{SCAR}} for fusion of FMs. We use fully connected network (FCN) and CNN as downstreams as used by previous research in related areas such as SER \cite{chetiaphukan23_interspeech} and shout intensity prediction \cite{fukumori2023investigating}. The CNN (Figure \ref{architecture} (b)) consists of a convolutional block with 1D CNN layer of 32 kernels with size 3 followed by maxpooling. The features are flattened and passed to a FCN block with two dense layers of 512 and 128 neurons followed by the output layer for binary classification. We use sigmoid as the activation function in the output layer. For FCN, we keep the same architectural details as the FCN block in CNN model. The CNN models trainable parameters ranges from 0.5 to 0.7M and for FCN, it ranges from 0.45 to 0.6M depending on the input representation dimension size. 


\subsection{\textbf{\texttt{SCAR}}}

The architecture of the proposed framework is illustrated in Figure \ref{architecture}. \textbf{\texttt{SCAR}} introduces a novel nested cross-attention mechanism that facilitates multi-stage interactions within the representational space of FMs, refining information exchange. Furthermore, a self-attention refinement module strengthens feature representations by emphasizing key cross-FMs cues while minimizing noise. The workflow of \textbf{\texttt{SCAR}} is as follows: First, representations are extracted from FMs and passed through a convolutional block with same modeling paradigm as done with individual FMs above. Following this, the flattened features are passed through the proposed nested cross-attention block that consists of two cross-attention layers sequentially. Let the flattened features of two FMs a and b be \( \mathbf{Z}_a \) and \( \mathbf{Z}_b \) respectively. First, the query, key, and value matrices for the first cross-attention step are computed as follows: $ \mathbf{Q}_{a1} = \mathbf{Z}_a \mathbf{W}_{Q1}^a, \quad \mathbf{K}_{b1} = \mathbf{Z}_b \mathbf{W}_{K1}^b, \quad \mathbf{V}_{b1} = \mathbf{Z}_b \mathbf{W}_{V1}^b $ and $ \mathbf{Q}_{b1} = \mathbf{Z}_b \mathbf{W}_{Q1}^b, \quad \mathbf{K}_{a1} = \mathbf{Z}_a \mathbf{W}_{K1}^a, \quad \mathbf{V}_{a1} = \mathbf{Z}_a \mathbf{W}_{V1}^a $. 

\noindent Using these, the cross-attention outputs for the first cross-attention step are computed as:

\[
\mathbf{Z}_a^{(1)} = \text{softmax} \left( \frac{\mathbf{Q}_{a1} \mathbf{K}_{b1}^T}{\sqrt{d_k}} \right) \mathbf{V}_{b1}
\]

\[
\mathbf{Z}_b^{(1)} = \text{softmax} \left( \frac{\mathbf{Q}_{b1} \mathbf{K}_{a1}^T}{\sqrt{d_k}} \right) \mathbf{V}_{a1}
\]

\noindent A second cross-attention step refines these representations: $
\mathbf{Q}_{a2} = \mathbf{Z}_a^{(1)} \mathbf{W}_{Q2}^a, \quad \mathbf{K}_{b2} = \mathbf{Z}_b^{(1)} \mathbf{W}_{K2}^b, \quad \mathbf{V}_{b2} = \mathbf{Z}_b^{(1)} \mathbf{W}_{V2}^b $ and $ \mathbf{Q}_{b2} = \mathbf{Z}_b^{(1)} \mathbf{W}_{Q2}^b, \quad \mathbf{K}_{a2} = \mathbf{Z}_a^{(1)} \mathbf{W}_{K2}^a, \quad \mathbf{V}_{a2} = \mathbf{Z}_a^{(1)} \mathbf{W}_{V2}^a $. The cross-attention outputs for the second cross-attention step are computed as follows:

\[
\mathbf{Z}_a^{(2)} = \text{softmax} \left( \frac{\mathbf{Q}_{a2} \mathbf{K}_{b2}^T}{\sqrt{d_k}} \right) \mathbf{V}_{b2}
\]

\[
\mathbf{Z}_b^{(2)} = \text{softmax} \left( \frac{\mathbf{Q}_{b2} \mathbf{K}_{a2}^T}{\sqrt{d_k}} \right) \mathbf{V}_{a2}
\]

Finally, a self-attention refinement step is applied to each modality separately:

\[
\mathbf{Z}_a^{(3)} = \text{softmax} \left( \frac{\mathbf{Z}_a^{(2)} \mathbf{Z}_a^{(2)T}}{\sqrt{d_k}} \right) \mathbf{Z}_a^{(2)}
\]

\[
\mathbf{Z}_b^{(3)} = \text{softmax} \left( \frac{\mathbf{Z}_b^{(2)} \mathbf{Z}_b^{(2)T}}{\sqrt{d_k}} \right) \mathbf{Z}_b^{(2)}
\]

\noindent This refinement module strengthens feature representations by highlighting essential cross-FM cues while reducing noise. The matrices \( \mathbf{Q}, \mathbf{K}, \mathbf{V} \) represent the query, key, and value matrices, while \( \mathbf{W}_Q, \mathbf{W}_K, \mathbf{W}_V \) are learnable weight matrices. The term \( d_k \) denotes the dimensionality of the key vectors used for scaling. The intermediate representations \( \mathbf{Z}_a^{(i)} \) and \( \mathbf{Z}_b^{(i)} \) denote the outputs at different stages of attention processing. This hierarchical cross-attention attention structure enables structured and progressive alignment of representations, enhancing the interaction between FMs while refining feature representations. The final fused representation is obtained by concatenating the refined outputs: $ \mathbf{Z}_{\text{fused}} = \text{Concat}(\mathbf{Z}_a^{(3)}, \mathbf{Z}_b^{(3)}) $. The concatenated features are then passed to a FCN block with two dense layers of 512 and 128 neurons respectively followed by the final output layer for binary classification. We keep the number of attention heads in nested cross attention for both the cross-attention stages as 2. For self-attention refinement stage, we keep the number of heads also as 2. \textbf{SCAR} consists of 1.8 to 2.3M trainable parameters depending on the dimension size of the FMs representations.


\begin{table}[hbt!]
\centering
\setlength{\tabcolsep}{15pt} 
\begin{tabular}{l|c|c|c|c}
\toprule
             & \multicolumn{2}{c|}{\textbf{E}} & \multicolumn{2}{c}{\textbf{C}} \\ 
\cmidrule(lr){2-3} \cmidrule(lr){4-5} 
\textbf{FM} & \textbf{FCN}  & \textbf{CNN}  & \textbf{FCN}  & \textbf{CNN}  \\ 
\midrule
U   & \cellcolor{wine10}2.43 & \cellcolor{wine10}2.21 & \cellcolor{wine5}7.68  & \cellcolor{wine5}7.35 \\
W2  & \cellcolor{wine10}2.51 & \cellcolor{wine10}2.24 & \cellcolor{wine10}4.36 & \cellcolor{wine10}3.29 \\
W   & \cellcolor{wine5}5.67  & \cellcolor{wine5}5.25  & \cellcolor{wine5}8.78  & \cellcolor{wine10}4.91 \\
Hu  & \cellcolor{wine5}3.62 & \cellcolor{wine10}2.93 & \cellcolor{wine5}10.94 & \cellcolor{wine5}10.23 \\
Wh  & \cellcolor{wine10}2.31 & \cellcolor{wine10}2.04 & \cellcolor{wine10}3.25 & \cellcolor{wine10}3.11 \\
\textbf{LB}  & \cellcolor{wine15}\textbf{1.64} & \cellcolor{wine15}\textbf{1.32} & \cellcolor{wine15}\textbf{2.19} & \cellcolor{wine15}\textbf{1.89} \\
\textbf{IB}  & \cellcolor{wine15}\textbf{1.87} & \cellcolor{wine15}\textbf{1.36} & \cellcolor{wine15}\textbf{2.10} & \cellcolor{wine15}\textbf{1.81} \\
\bottomrule
\end{tabular}
\caption{EER scores are given in \% for various FMs; E, C stands for English and Chinese subset; FMs Abbreviations are given as follows: Unispeech-SAT -- U, Wav2vec2 -- W2, WavLM -- W, Whisper -- Wh, XLS-R -- X, x-vector -- XV, MMS -- MM, HuBERT -- Hu, LB -- LanguageBind, IB -- ImageBind; The intensity represents performance levels, with darker shades signifying lower EER values and lighter shades indicating higher EER values; The abbreviations of the FMs and the intensity scale in this table are maintained consistently for Table \ref{tab:2}}
\label{tab:1}
\end{table}

\begin{table}[hbt!]
\centering
\setlength{\tabcolsep}{13pt} 
\renewcommand{\arraystretch}{1.0} 
\adjustbox{max width=\textwidth}{
\begin{tabular}{l|c|l|l|l}
\toprule
\multicolumn{1}{c|}{\multirow{2}{*}{Combinations}} & \multicolumn{2}{c|}{Concatenation} & \multicolumn{2}{c}{\textbf{\texttt{SCAR}}} \\
\cmidrule(lr){2-3} \cmidrule(lr){4-5}
 & \textbf{E} & \textbf{C} & \textbf{E} & \textbf{C} \\
\midrule
U + W2   & \cellcolor{wine5}3.93 & \cellcolor{wine5}6.63 & \cellcolor{wine10}2.45 & \cellcolor{wine5}4.05 \\
U + W    & \cellcolor{wine5}3.48 & \cellcolor{wine5}7.05 & \cellcolor{wine5}3.61 & \cellcolor{wine5}6.56 \\
U + Hu   & \cellcolor{wine5}3.06 & \cellcolor{wine5}5.21 & \cellcolor{wine10}2.96 & \cellcolor{wine5}4.62 \\
U + Wh   & \cellcolor{wine5}4.11 & \cellcolor{wine5}4.35 & \cellcolor{wine10}2.12 & \cellcolor{wine10}2.84 \\
U + LB   & \cellcolor{wine5}3.42 & \cellcolor{wine10}2.94 & \cellcolor{wine10}1.95 & \cellcolor{wine10}2.45 \\
U + IB   & \cellcolor{wine10}2.90 & \cellcolor{wine5}3.43 & \cellcolor{wine10}1.89 & \cellcolor{wine10}1.97 \\
\midrule
W2 + W   & \cellcolor{wine10}2.65 & \cellcolor{wine5}5.14 & \cellcolor{wine10}2.13 & \cellcolor{wine5}3.94 \\
W2 + Hu  & \cellcolor{wine5}3.04 & \cellcolor{wine5}5.87 & \cellcolor{wine10}2.96 & \cellcolor{wine5}5.31 \\
W2 + Wh  & \cellcolor{wine5}3.03 & \cellcolor{wine5}6.11 & \cellcolor{wine10}2.63 & \cellcolor{wine5}3.07 \\
W2 + LB  & \cellcolor{wine10}2.07 & \cellcolor{wine10}2.93 & \cellcolor{wine10}1.92 & \cellcolor{wine10}1.10 \\
W2 + IB  & \cellcolor{wine10}2.35 & \cellcolor{wine5}4.94 & \cellcolor{wine10}1.85 & \cellcolor{wine5}2.04 \\
\midrule
W + Hu   & \cellcolor{wine5}3.01 & \cellcolor{wine5}4.34 & \cellcolor{wine10}2.86 & \cellcolor{wine5}3.57 \\
W + Wh   & \cellcolor{wine5}3.76 & \cellcolor{wine5}4.06 & \cellcolor{wine5}3.63 & \cellcolor{wine10}2.92 \\
W + LB   & \cellcolor{wine10}2.98 & \cellcolor{wine5}3.92 & \cellcolor{wine10}1.95 & \cellcolor{wine5}2.94 \\
W + IB   & \cellcolor{wine10}2.93 & \cellcolor{wine5}3.37 & \cellcolor{wine10}2.13 & \cellcolor{wine5}2.45 \\
\midrule
Hu + Wh  & \cellcolor{wine5}3.21 & \cellcolor{wine5}4.86 & \cellcolor{wine10}2.65 & \cellcolor{wine5}3.02 \\
Hu + LB  & \cellcolor{wine5}4.04 & \cellcolor{wine5}6.02 & \cellcolor{wine5}3.02 & \cellcolor{wine10}2.93 \\
Hu + IB  & \cellcolor{wine5}3.03 & \cellcolor{wine5}5.07 & \cellcolor{wine10}2.73 & \cellcolor{wine5}4.90 \\
\midrule
Wh + LB  & \cellcolor{wine10}1.38 & \cellcolor{wine10}1.89 & \cellcolor{wine10}1.24 & \cellcolor{wine10}1.66 \\
Wh + IB  & \cellcolor{wine10}2.67 & \cellcolor{wine10}1.93 & \cellcolor{wine10}1.53 & \cellcolor{wine10}1.81 \\
\midrule
\textbf{LB + IB}  & \cellcolor{wine15}\textbf{1.21} & \cellcolor{wine15}\textbf{1.43} & \cellcolor{wine15}\textbf{1.15} & \cellcolor{wine15}\textbf{1.02} \\
\bottomrule
\end{tabular}}
\caption{Evaluation scores (EER in \%) for pairwise combinations of FMs; E, C represents the English and Chinese subsets}
\label{tab:2}
\end{table}

\begin{figure}[!bt]
    \centering
    \subfloat[]{%
        \includegraphics[width=0.2\textwidth]{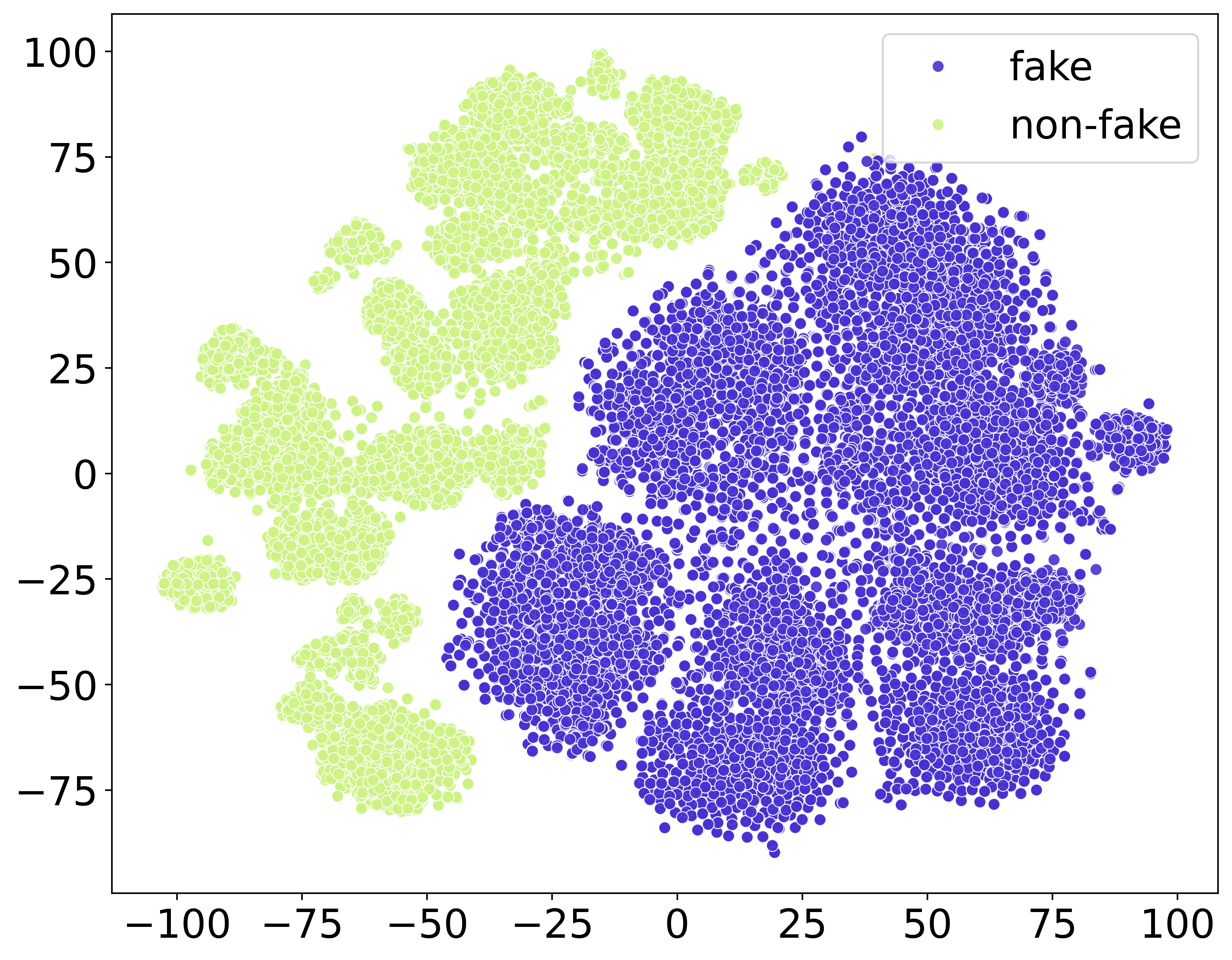}
    }
    \hspace{0.01\textwidth}
    \subfloat[]{%
        \includegraphics[width=0.2\textwidth]{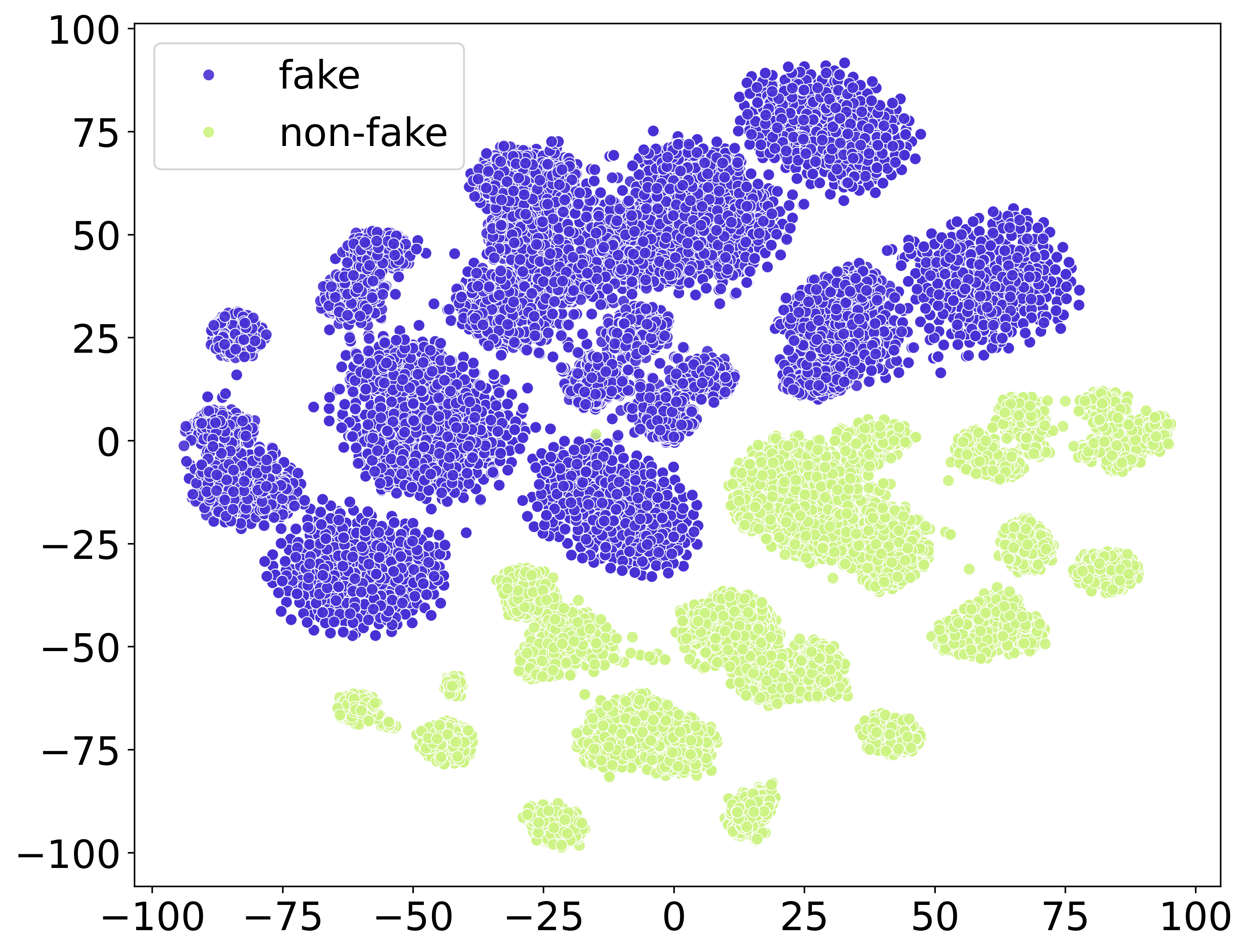}
    }
    \par\bigskip 
    \subfloat[]{%
        \includegraphics[width=0.2\textwidth]{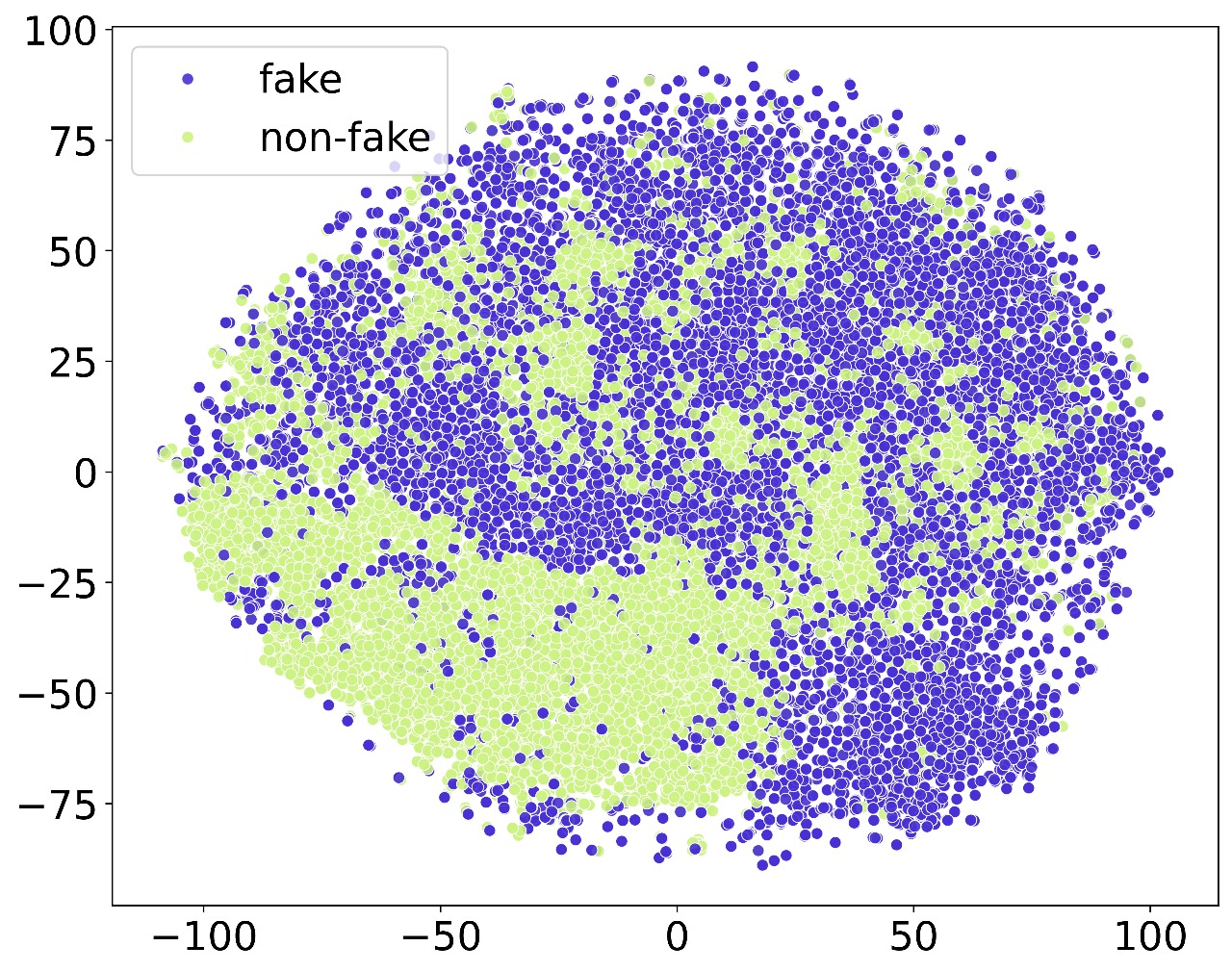}
    }
    \hspace{0.01\textwidth}
    \subfloat[]{%
        \includegraphics[width=0.2\textwidth]{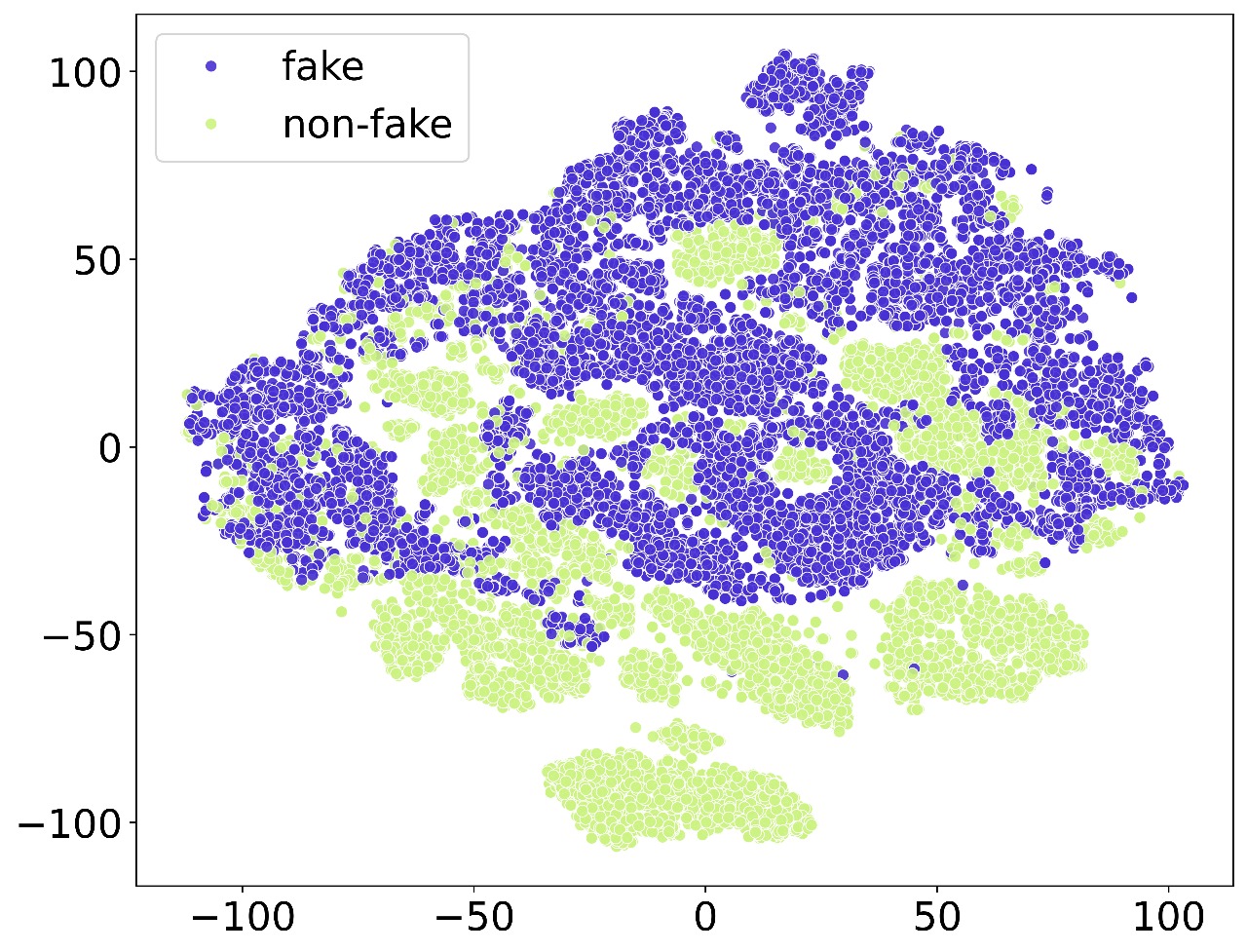}
    }
    \caption{The subfigures (a) LB (b) IB (c) Wav2vec2 and (d) Whisper represent the t-SNE plot visualization}
    \label{tsne}
\end{figure}

\section{Result and Analysis}

\subsection{Dataset}

\noindent We utilize the sole open-source dataset for EFD, introduced by Zhao et al. \cite{zhao2024emofake}. This dataset includes recordings in both English and Chinese, containing samples from multiple speakers across five primary emotions: Neutral, Happy, Angry, Sad, and Surprise. Fake emotional expressions were created using seven open-source Emotional Voice Conversion models, which modify emotional cues while preserving speaker identity. The dataset is organized into separate English and Chinese subsets, each with designated training, development, and test splits. Both the English and Chinese partitions consist of 27,300 training samples, 9,100 development samples, and 17,500 test samples. We use the official split for training, validation, and evaluation of the models.

\noindent \textbf{Training Details}: We use the Adam as the optimizer to train the models with a learning rate of 1e-3 and a batch size of 32. For the loss function, we use binary cross-entropy. To avoid overfitting, we use dropout and early stopping.

\subsection{Experimental Results}
We use Equal Error Rate (EER) as the evaluation metric following previous research on EFD \cite{zhao2024emofake}. Table \ref{tab:1} presents the evaluation scores of downstream models trained on representations from various FMs. We observe that MFMs consistently achieve lower EER values across both language subsets and downstream network architectures in comparison to unimodal AFMs. \textit{This validates our hypothesis that MFMs outperform AFMs in EFD as their cross-modal pretraining allows them to learn emotional patterns from multiple modalities. This enhances their ability to detect unnatural emotional shifts and inconsistencies in manipulated audio.} Among the AFMs, Whisper achieved the highest performance, which can be attributed to its multilingual pretraining. This aligns with previous research on synthetic speech detection \cite{chetia-phukan-etal-2024-heterogeneity}, which found that multilingual AFMs better capture pitch, tone, and intensity variations, enhancing their ability to detect synthetic or manipulated speech. Overall, CNN-based models generally yield lower EER scores compared to FCNs with all the FMs. We also plot the t-SNE plots of raw representations from the FMs in Figure \ref{tsne}. We can observe clear clusters across the classes for MFMs and thus providing support to our results. \par

Table \ref{tab:2} presents the evaluation scores for different combinations of FMs. We use concatenation as baseline fusion technique. For modeling concatentation-based fusion, we follow similar architecture as \textbf{\texttt{SCAR}} and discard nested cross-attention block and the self-attention refinement. We follow the same training specifications as used for \textbf{\texttt{SCAR}}. We can observe clear dominance of \textbf{\texttt{SCAR}} based fusions of FMs compared to concatenation-based technique and this shows the strength of \textbf{\texttt{SCAR}} for effective fusion. The topmost performance is achieved by fusion of MFMs i.e. LB and IB through \textbf{\texttt{SCAR}} and this shows the emergence of strong complementary behavior among them.


\noindent \textbf{Comparison to SOTA:} We compare our best model \textbf{\texttt{SCAR}} with LB and IB with previous SOTA work \cite{zhao2024emofake}. They reported the EER values as 3.65\% and 8.34\% for English and Chinese subsets respectively. However, \textbf{\texttt{SCAR}} with LB and IB reported the best EER of 1.15\% and 1.02\% for English and Chinese respectively and thus setting new SOTA for EFD.

\section{Conclusion}
Our study shows that MFMs outperform AFMs for EFD due to their cross-modal pretraining, which enables better recognition of unnatural emotional shifts in manipulated audio. Further, we also introduce, \texttt{\textbf{SCAR}}, a novel framework for fusion of FMs that leverages nested cross-attention and self-attention refinement. \texttt{\textbf{SCAR}} with the fusion of MFMs achieves achieves SOTA results, surpassing standalone FMs and baseline fusion method, establishing a new benchmark for EFD. Also,  our study serves as a guide for future research exploring FMs for EFD.

\printbibliography

\end{document}